\documentclass[letterpaper, 10 pt, conference]{ieeeconf}  

\IEEEoverridecommandlockouts                              

\usepackage{epsfig} 
\usepackage{amsmath} 
\usepackage{amssymb}  
\usepackage{amsthm}
\usepackage{booktabs}
\usepackage{mathrsfs}

\newtheorem{thm}{Theorem}
\newtheorem{lem}{Lemma}

\theoremstyle{definition}

\newtheorem{ass}{Assumption}
\theoremstyle{remark}
\newtheorem{rem}{Remark}
\usepackage{xcolor}

\title{\LARGE \bf
Robust Control Barrier Functions for Safe Control Under Uncertainty Using Extended State Observer and Output Measurement}

\author{Jinfeng Chen, Zhiqiang Gao,  Qin Lin
\thanks{$^{}$All authors are with the Center for Advanced Control Technologies (CACT), Cleveland State University. Corresponding author: Qin Lin
{\tt\small q.lin80@csuohio.edu}}%
\thanks{This material is based upon work supported by the National Science Foundation under Grants No. 2301543.}
}

\begin{document}

\maketitle
\thispagestyle{empty}
\pagestyle{empty}

\begin{abstract}
Control barrier functions-based quadratic programming (CBF-QP) is gaining popularity as an effective controller synthesis tool for safe control. However, the provable safety is established on an accurate dynamic model and access to all states. To address such a limitation, this paper proposes a novel design combining an extended state observer (ESO) with a CBF for safe control of a system with model uncertainty and external disturbances only using output measurement. Our approach provides a less conservative estimation error bound than other disturbance observer-based CBFs. Moreover, only output measurements are needed to estimate the disturbances instead of access to the full state. The bounds of state estimation error and disturbance estimation error are obtained in a unified manner and then used for robust safe control under uncertainty. We validate our approach’s efficacy in simulations of an adaptive cruise control system and a Segway self-balancing scooter.

\end{abstract}


\section{Introduction}

Safety is critical in controller design for robotic systems such as self-driving cars, aerial vehicles, and industrial robots that operate in dynamic and uncertain environments. Control barrier functions combined with quadratic programming (CBF-QP) method is a novel controller synthesis method that solves an optimization problem online with safety as a hard constraint \cite{ames2016control}. Combined with a control Lyapunov function (CLF), as a dual concept of CBF, stability can also be included to form a CBF-CLF-QP. 

CBF-QP renders a forward invariant set to enforce a dynamic system's trajectory stay in a safe set over an infinite time horizon. However, the significant limitations include: 1) The assurance of safety relies on precisely modeled dynamics. Uncertainties due to parametric error, unmodeled dynamics, external disturbance, etc. hinder the safe deployment of robotic systems using CBF-QP. 2) Existing CBF-QP works assume access to full state, which is not practical. 



There are four main categories of related works: 1) \emph{input-to-state robust CBF-QP}: In \cite{alan2021safe}, the authors did not estimate the disturbance but instead used the disturbance bound for robust control. All states were assumed to be available. 2) \emph{learning-based adaptive CBF-QP}: Machine learning has been leveraged to compensate for model errors in CBF-QP, such as reinforcement learning \cite{choi2020reinforcement}, episodic learning \cite{taylor2020learning}, and online learning \cite{Emanuel2022}. The fundamental limitation is the lack of formal safety guarantees. 3) \emph{observer-based CBF}: \cite{zhao2020adaptive} used $\mathcal{L}1$ adaptive control to estimate disturbance. \cite{dacs2022robust} and \cite{alan2023disturbance} used disturbance observers (DOBs) to estimate disturbances in the dynamics of CBFs. These works all assumed available state information. 4) \emph{measurement-robust CBF}:  In \cite{dean2021guaranteeing}, the authors used machine learning to establish the mapping from the perception to the states. However, the dynamics uncertainty was not considered. In \cite{agrawal2022safe}, the authors used an Input-to-State Stable observer and a ``Bounded Error" observer to estimate the states for robust safe control. However, the disturbances were not estimated and compensated. To sum up, \textbf{there is no existing work simultaneously dealing with state estimation and disturbance estimation for a robust safe control using CBF-QP.}

 To close such a gap, we propose a novel design combining ESO with CBF for robust safe control. Our approach is named as ESOR-QP. First, owing to the intrinsic feature of ESO, we are able to estimate the system states and total disturbance including \emph{internal disturbance} (i.e., unknown or unmodelled parts of the plant dynamics) and \emph{external disturbance} (i.e., various perturbations from the outside but affecting the evolution of dynamics) in a unified observer. Second, the error bound of state estimation and disturbance estimation is derived for a robust optimization to empower a robust safe control, i.e., the safety is still guaranteed in the presence of disturbance and state estimation errors.


The contributions of our work are summarized as follows:
\begin{itemize}

\item To the best of our knowledge, it is the first-of-its-kind safe controller that combines ESO and CBF-QP guaranteeing safety in the presence of internal and external disturbances. 

\item Compared with state-of-the-art observer-based CBF-QP, we estimate state and disturbance simultaneously in a unified framework based on ESO. The theoretical estimation error bound is less conservative than the existing DOB-based CBF-QP. 
\end{itemize}

The remainder of this paper is structured as follows. We go through necessary preliminary in Sec. \ref{sec:background}. The formulation of ESOR-QP is in Sec. \ref{sec:ESO-CLF-CBF-QP}. The simulation results are in Sec. \ref{sec:simulation}. The concluding remarks and future work are in Sec. \ref{sec:conclusion}.


\section{PRELIMINARY}
\label{sec:background}

\subsection{Notation}

\(\mathbb{R}^+\), \(\mathbb{R}^n\), and \(\mathbb{R}^{n\times m}\) denote sets of non-negative real numbers, \(n\)-dimensional real vectors, and \(n\) by \(m\) dimensional real matrices, respectively. \(\mathbb{I}_n\) represents an \(n\) by \(n\) identity matrix. \(\|\cdot\|\) denotes \(2\)-norm of a vector (or a matrix). \(\|\cdot\|_{\mathcal{L}_{1}}\) and \(\|\cdot\|_{\mathcal{L_\infty}}\) denote \(\mathcal{L}_1\)-norm and \(\mathcal{L}_{\infty}\)-norm of a piecewise-continuous integrable function, respectively. 

\subsection{Model and Coordinate Transformation}
Consider the following multi-input multi-output (MIMO) nominal nonlinear control-affine system
\begin{equation}\label{eq_1}
    \begin{cases}
        \dot x =f(x)+g(x)u \\
        y=\zeta(x)
    \end{cases}
\end{equation}
where $x\in X\subset \mathbb{R}^n$, $u\in U \subset \mathbb{R}^m$, and $y\in\mathbb{R}^m$ are the state, input, output vectors, respectively. \(f:\mathbb{R}^n \rightarrow \mathbb{R}^n\), \(g:\mathbb{R}^n\rightarrow \mathbb{R}^{n\times m}\), and \(\zeta:\mathbb{R}^n\rightarrow\mathbb{R}^m\) are known and smooth mappings. Suppose \(X\) and \(U\) are compact sets. In the framework of the conventional ESO, the model uncertainty and external disturbances are lumped together and placed in the control input channel. In this paper, the total disturbances corresponding to all measurements are added to the system after coordinate transformation, which will be discussed soon. If the system (\ref{eq_1}) has a well-defined \emph{vector relative degree}, the theory developed to convert a nonlinear control-affine system to a normal form for single-input single-output (SISO) systems can be extended to MIMO systems (\ref{eq_1}) \cite{isidori1985nonlinear}. Moreover, the strict assumption of an existed vector relative degree has been relaxed to \emph{invertibility} (i.e., the system has a nonsingular transfer function) via a \emph{dynamic extension} \cite{wu2020performance}.



\begin{ass}\label{ass1}
    The system (\ref{eq_1}) is invertible.
\end{ass}

Since a MIMO system can be divided into several MISO subsystems under Assumption \ref{ass1} \cite{wu2020performance}, we can consider the following SISO nonlinear affine system for simplicity
\begin{equation}\label{eq_2}
    \begin{cases}
        \dot x=f(x)+g_i(x)u_i\\
        y_i=\zeta_i(x)
    \end{cases}
\end{equation}
where \(x\in\mathbb{R}^r\), \(u_i\in \mathbb{R}\) and \(y_i\in \mathbb{R}\) are the \(i\)-th input and output, respectively, \(f(x)\) and \(g_i(x)\) are \(r\)-dimensional vectors of functions of the state \(x\), and \(\zeta_i(x)\) is a scalar function. If the relative degree of the system (\ref{eq_2}) is \(r\), then\footnote{\(L_f \zeta_i(x)\triangleq\frac{\partial \zeta_i(x)}{\partial x}f(x)\), \(L_{g_i} L_f \zeta_i(x)\triangleq\frac{\partial(L_f \zeta_i)}{\partial x}g_i(x)\), \(L_f^k\zeta_i(x)\triangleq\frac{\partial(L_f^{k-1}\zeta_i)}{\partial x}f(x)\).}

(i) \(L_{g_i}L_f^k \zeta_i(x)=0\) for all \(k < r -1\)
and for all \(x\) in a neighborhood of a certain point \(x^\circ\), 

(ii) \(L_{g_i} L_f^{r-1} \zeta_i(x^\circ) \neq 0\). 

The coordinate transformation matrix \(z=\varPhi (x)=[\zeta_i(x), L_f \zeta_i(x), \cdots, L_f^{r-1}\zeta_i(x)]^T\)can convert the nonlinear affine system (\ref{eq_2}) into the following normal form
\begin{equation}\label{eq_5}
    \begin{cases}
        \dot{z}_1 =z_2 \\
        \dot{z}_2 =z_3\\
        \quad \  \vdots\\
        \dot{z}_r =b(z)+a(z)u_i+d_i(t,z,u)
    \end{cases}
\end{equation}
where \(z=[z_1, z_2, \cdots, z_r]^T\), \(a(z)=L_{g_i}L_f^{r-1}\zeta_i(\varPhi^{-1}(z))\), and \(b(z)=L_f^r \zeta_i(\varPhi^{-1}(z))\). Note that the term \(d_i(t,z,u)\) is added in \eqref{eq_5} as a total disturbance due to various uncertainties affecting the specific single output measurement \(y_i\) \cite{chen2020active}. Since the system (\ref{eq_5}) is in the form of an integrator chain, an ESO can be used to estimate the state \(z\) and the total disturbance \(d_i(t,z,u)\). 
Therefore, for each measured output, a similar ESO can be designed \cite{wu2020performance}. Hereafter, we interchangeably represent \(d_i(t,x,u)\) as \(d_i\) or \(d_i(t)\), because \(d_i(t)\), in this paper, is considered to be an unknown signal and its relation to \(x\) and \(u\) is not available. 



\subsection{Control Barrier Functions}
The converted nonlinear system (\ref{eq_5}) can still be written in a nonlinear affine form as (\ref{eq_1}) with the total disturbances added. Without loss of generality, the following discussion about CBF-QP is based on a nonlinear affine form given by
\begin{equation}\label{eq_affine_d}
    \dot{x}=f(x)+g(x)u+d, 
\end{equation}
where \(d\in \mathbb{R}^n\) includes all total disturbances corresponding to all measured outputs.

The definition of CBF in the context of uncertainty is given in \cite{zhao2020adaptive}. A continuous and differentiable function \(h: \mathbb{R}^n \rightarrow \mathbb{R}\) is called a CBF for the system \eqref{eq_affine_d}, if 
\begin{equation}\label{eq_8}
    \sup\limits_{u\in U} \left(L_{f}h(x)+L_g h(x)u+h_x(x)d\right)\geq -\beta(h(x))
\end{equation}
for all \(t\geq 0\) and \(x\in X\), where \(h_x(x)= \frac{\partial h(x)}{\partial x}\), and \(\beta(s)\) is an extended class \(\mathcal{K}\) function.
In particular, we usually choose \(\beta(s)=\gamma s\) for a constant \(\gamma>0\). From \eqref{eq_8}, we have a set of control signals ensuring that the states of the system \eqref{eq_affine_d} are always in a safe set \(\mathcal{C}\) if \(h_x(x)\neq 0\) for all \(x\in \partial \mathcal{C}\), and the initial states are in the safe set:
\begin{equation}\label{eq_10}
    \begin{array}{r@{}l}
        K_{\text{cbf}}(t,x,d)\triangleq \{u\in U: L_{f}h(x)+ & L_g h(x)u \\
        +h_x(x)d & \geq -\gamma h(x)\} 
    \end{array}
\end{equation}
where \(\partial\mathcal{C}\) represents the boundary of \(\mathcal{C}\). 

\subsection{QP-CBF Formulation}
In order to combine CBF to guarantee safety, the control problem is formulated as a quadratic program with a CBF as a hard constraint. The QP formulation is as follows \cite{ames2016control}:

\begin{equation}
\begin{split}
    u^* (x) & = \operatorname*{arg\,min}_{u\in U} \|u-k(x)\|^2 \\
    \text{s.t.} 
    \quad L_{f}h(x) & +L_g h(x)u +h_x(x)d+\beta (h(x)) >0
    \label{eq_11}
\end{split}
\end{equation}
where \(k(x)\) is a nominal control law. 

\section{ESO-based Robust QP Control with CBF}
\label{sec:ESO-CLF-CBF-QP}

\subsection{ESO and Estimation Error Bound}
In practice, we only have output measurements $y$. To implement CBF-QP, the state $x$ and disturbance $d$ in (\ref{eq_11}) need to be estimated by ESOs. After converting the original nonlinear control-affine system into a normal form and adding the corresponding disturbance for each measurement, we use $x$ to represent state $z$ in (\ref{eq_5}) just for symbolic consistence for CBF. Equation (\ref{eq_5}) can be written in a matrix form:
\begin{equation}\label{eq_add}
    \begin{cases}
        \dot{x}=A_0 x+B_0 (b(x)+a(x)u_i +d_i)\\
        y_i=C_0 x
    \end{cases}
\end{equation}
where \(A_0=\begin{bmatrix}
    0 & 1 & \cdots & 0\\
    \vdots & \vdots & \ddots & \vdots\\
    0 & 0 & \cdots & 1\\
    0 & 0 & \cdots & 0
\end{bmatrix}_{r\times r}\), 
\(B_0=\begin{bmatrix}
    0\\
    \vdots\\
    0\\
    1
\end{bmatrix}_{r\times 1}\),
\(C_0=[1, 0, \cdots, 0]_{1\times r}\),
\(r\) is the relative degree of $y_i$ with respect to $u_i$ for $i\in [1, m]$, 
$d_i$ is the corresponding disturbance for $y_i$, 
\(a(x)=L_{g_i}L_f^{r-1}\zeta_i(\varPhi^{-1}(x))\), and \(b(x)=L_f^r \zeta_i(\varPhi^{-1}(x))\).

Since the disturbance for each measurement can be treated as an extended state \cite{gao2003scaling}, the augmented system becomes: 
\begin{equation}\label{eq_13}
    \begin{cases}
        \begin{bmatrix}
            \dot{x}\\
            \dot{f}_i
        \end{bmatrix} = A \begin{bmatrix}
            x\\
            f_i
        \end{bmatrix} +B(b(\hat{x})+a(\hat{x})u_i)+D\dot{f}_i \\
        y_i=C \begin{bmatrix}
            x &
            f_i
        \end{bmatrix}^T
    \end{cases}
\end{equation}
where \(A=\begin{bmatrix}
    A_0 & B_0\\
    0_{1\times r} & 0
\end{bmatrix}_{(r+1)\times(r+1)}\), 
\(B=\begin{bmatrix}
    B_0\\
    0
\end{bmatrix}_{(r+1)\times 1}\), 
\(C=[1, 0, \cdots, 0]_{1\times (r+1)}\), 
\(D=[0, \cdots, 0, 1]_{1\times(r+1)}^T\), 
\(\hat{x}\) is the state of the ESO in (\ref{eq_14}), and the actual total disturbance is \(f_i=b(x)-b(\hat{x})+(a(x)-a(\hat{x}))u_i+d_i\) rather than \(d_i\). 

An ESO is designed upon the formulation in (\ref{eq_13}):
\begin{equation}\label{eq_14}
    \begin{bmatrix}
        \dot{\hat{x}} \\
        \dot{\hat{f}}_i
    \end{bmatrix} = A \begin{bmatrix}
        \hat{x}\\
        \hat{f}_i
    \end{bmatrix} + B (b(\hat{x}) +a(\hat{x})u_i)+L\left( y_i-C \begin{bmatrix}
        \hat{x}\\
        \hat{f}_i
    \end{bmatrix}\right)
\end{equation}
where \(\hat{x}\) is the estimation of the state \(x\), \(\hat{f}_i\) is the estimation of the disturbance \(f_i\), \(L\in \mathbb{R}^{r+1}\) is called \emph{observer gain}. Due to the special structure of \((A,C)\), the system (\ref{eq_13}) is observable and the eigenvalues of \(A-LC\) can be placed at \(-\omega_o\), where \(\omega_o\) is called \emph{observer bandwidth} \cite{gao2003scaling}. 

A discrete version of ESOR-QP will be introduced for the following reasons: 1) it is a supplement to the aforementioned continuous formulation to deliver a comprehensive introduction of our proposed method; 2) it facilitates the comparison with the adaptive robust QP control (aR-QP) in \cite{zhao2020adaptive}, which also uses a discrete formulation; 3) the theoretical estimation error bound can be leveraged from our recent work \cite{chen2022relationship} based on such a discrete formulation. Note that in the remainder of this section, we will show that \textbf{such an estimation error bound of disturbance can also be directly used in a continuous system without discretization}.

The continuous system (\ref{eq_add}) can be discretized as:
\begin{equation}\label{eq_15}
\small
    \begin{cases}
        x(k+1)=A_0 x(k)+B_0 (b(x(k))+a(x(k))u_i(k) + d_i(k))\\
        y_i(k)=C_0 x(k)
    \end{cases}
\end{equation}
where \(a(x(k))\) and \(b(x(k))\) are the discretizations of \(a(x)\) and \(b(x)\). After including the disturbance for \(y_i(k)\) as an extended state, the system (\ref{eq_15}) can be augmented into:
\begin{equation}\label{eq_16}
    \begin{split}
        \begin{bmatrix}
            x(k+1)\\
            f_i(k+1)
        \end{bmatrix}& = A\begin{bmatrix}
            x(k)\\
            f_i(k)
        \end{bmatrix}+B(b(\hat{x}(k)) \\
        & +a(\hat{x}(k))u_i(k)) + D\Delta f_i(k)\\
        y_i(k) & =C\begin{bmatrix}
            x(k) & f_i(k)
        \end{bmatrix}^T
    \end{split}
\end{equation}
where \(A=\begin{bmatrix}
    A_0 & B_0\\
    0_{1\times r} & 1
\end{bmatrix}_{(r+1)\times(r+1)}\), 
\(B\), \(C\), and \(D\) are the same as those in \eqref{eq_13}, 
\(\hat{x}(k)\) is the state vector of the ESO in (\ref{eq_17}), the actual total disturbance is \(f_i(k)=b(x(k))-b(\hat{x}(k))+(a(x(k))-a(\hat{x}(k)))u_i(k)+d_i(k)\) rather than \(d_i(k)\), and \(\Delta f_i(k)=f_i(k+1)-f_i(k)\).

Similar to the continuous system, a discrete ESO can be used to estimate the state and the disturbance as follows:
\begin{equation}\label{eq_17}
    \begin{array}{r@{}l}
        \begin{bmatrix}
            \hat{x}(k+1)\\
            \hat{f}_i(k+1)
        \end{bmatrix} = A\begin{bmatrix}
            \hat{x}(k)\\
            \hat{f}_i(k)
        \end{bmatrix} +B(b(\hat{x}(k))+a(\hat{x}(k))u_i(k)) &  \\
        + L\left( y_i(k) - C \begin{bmatrix}
            \hat{x}(k)\\
            \hat{f}_i(k)
        \end{bmatrix} \right) & 
    \end{array}
\end{equation}
where \(\hat{x}(k)\) and \(\hat{f}_i(k)\) are the estimations of \(x(k)\) and \(f_i(k)\), and \(L\) is the observer gain. Since the system (\ref{eq_17}) is discrete, the eigenvalues of \(A-LC\) need to be placed at \(\omega_o\) inside the unit circle to make the observer converge. 

\begin{ass}\label{ass2}
    There exist positive known constants \(l_f\) and \(b_f\) such that for any \(x\in X\), \(u\in U\), and \(t\geq0\), the following inequality holds:
    \begin{equation}\label{eq_6}
        \left|\dfrac{\partial f_i(t,x,u)}{\partial t}\right|\leq l_f, \quad |f_i(t,x,u)|\leq b_f. 
    \end{equation}
\end{ass}

\begin{rem}
    This assumption essentially states that the change and the magnitude of the disturbance $f_i$, rather than \(d_i\), for each output measurement are bounded. 
\end{rem}

Based on the results of our previous work \cite{chen2022relationship}, the estimation error of \(\hat{f}_i\) is as follows:
\begin{equation}\label{eq_18}
    f_i(k)-\hat{f}_i(k)=p(k)* \Delta f_i(k)
\end{equation}

\begin{equation}\label{eq_19}
\footnotesize
    \begin{array}{l}
         p(k)=   \\
          \begin{cases}
                1 & 1\leq k \leq r+1\\
                \sum\limits_{i = 1}^{r+1}  \frac{1}{(i-1)!}(1-\omega_o)^{i-1}\left(\prod\limits_{j=-i+1}^{-1}(k+j)\right)\omega_o^{k-i} & k\ge r+2.
            \end{cases} 
    \end{array}
\end{equation}
For simplicity, let \(\prod_{j=0}^{-1}(k+j)=1\). From the definition of derivative and Assumption \ref{ass2}, we have \(f_i(t+T)-f_i(t)\leq l_f T\), where \(T\) is the estimation sample time, \(l_f\) is the upper bound defined in (\ref{eq_6}). 
For simplicity, let us define
\begin{equation}\label{eq_add2}
    \gamma(\omega_o, T)=\left(\sum\limits_{k=1}^{\infty}p(k)\right)l_f T
\end{equation}

\begin{lem}\label{lem1}
    Given the system (\ref{eq_15}) and the discrete ESO in (\ref{eq_17}), subject to Assumption \ref{ass2}, the estimation error bound of disturbance can be obtained as
    \begin{equation}\label{eq_20}
        |f_i(k)-\hat{f}_i(k)|\leq \gamma(\omega_o, T). 
    \end{equation}
\end{lem}

\begin{rem}
    Lemma \ref{lem1} implies that the estimation error of disturbance is related to the observer bandwidth \(\omega_o\) and the estimation sample time \(T\). If \(0\leq \omega_o<1\), the estimation error of disturbance approaches zero by reducing \(T\). However, \(T\) and \(\omega_o\) can not be reduced to unreasonably small values due to the measurement noise \cite{chen2022relationship}. 
\end{rem}
\begin{rem}
    The estimation error bound of disturbance can be directly used in continuous-time systems because (\ref{eq_18}) only relates to \(\omega_o\), \(r\), \(T\) and \(l_f\), and the estimation error bound for the same system stays same regardless of whether it is calculated in the continuous-time or discrete-time domain. To obtain an accurate error bound for continuous-time systems, the estimation sample time \(T\) should be small, such as \(0.1\) ms, and the observer bandwidth in the discrete-time domain needs to be converted via \(z=e^{sT}\).
\end{rem}

\begin{rem}
    Since the convergence of ESO has been proved in \cite{xue2015performance}, the accurate estimation error bound of disturbance in (\ref{eq_20}) converges.
\end{rem}

Note that the CBF-QP problem is formulated in a continuous-time domain, see (\ref{eq_11}). The estimation error of \(x\) also needs to be considered. Lemma \ref{lem2} considers the estimation error bounds in the continuous-time domain.
\begin{lem}\label{lem2}
Given the system (\ref{eq_add}) and the continuous ESO in (\ref{eq_14}), subject to Assumption \ref{ass2}, the estimation error bounds of states \(x\) and \(\dot{x}\) are
    \begin{align}
        \|x-\hat{x}\|_{\mathcal{L}_{\infty}} & \leq \|G(s)\|_{\mathcal{L}_1} B_0 \gamma(\omega_o, T) \label{eq_22}\\
        \|\dot{x}-\dot{\hat{x}}\|_{\mathcal{L}_{\infty}} & \leq \|H(s)\|_{\mathcal{L}_1} B_0 \gamma(\omega_o, T) \label{eq_22b}
    \end{align}
    where $G(s)=(s\mathbb{I}_{r}-(A_0-L_0C_0))^{-1}$, $H(s)=(A_0-L_0C_0)G(s)+\mathbb{I}_{r}$, 
    and $L_0$ is the first $n$ elements of $L$. 
\end{lem}

\begin{lem}\label{lem3}
    Given the Assumption \ref{ass2}, for any \(t>0\), \(x\in X\) and \(u_i\in U\), we have 
    \begin{equation}\label{eq_29}
        \|\dot{x}(t)\| \leq \phi, 
    \end{equation}
    where \(\phi\triangleq \max\limits_{x\in X, u_i\in U}\|f(x)+g(x)u_i\|+b_f\). 
\end{lem}

\subsection{Robust Controller Synthesis}
After converting to the normal form, (\ref{eq_add}) can be written in a nonlinear control-affine form as follows:
\begin{equation}\label{eq_30}
    \dot{x}=f(x)+g(x)u_i+B_0 d_i
\end{equation}
where \(f(x)=A_0 x+B_0 b(x)\) and \(g(x)=B_0 a(x)\). 
The derivative of CBF with respect to \(t\) can be written as
\begin{equation}\label{eq_31}
    \begin{split}
        \dot{h} = & L_fh(\hat{x})+L_gh(\hat{x})u_i+h_{x}(\hat{x})B_0\hat{f}_i+h_x(x)(\dot{x}-\dot{\hat{x}})\\
          & +(h_x(x)-h_{x}(\hat{x}))\dot{\hat{x}}+h_x(\hat{x})L_0C_0(x-\hat{x}). 
    \end{split}
\end{equation}

\begin{equation}\label{eq_33}
    \begin{split}
        \Psi_h(t,\hat{x},u) & \triangleq L_fh(\hat{x})+L_gh(\hat{x})u_i+h_{x}(\hat{x})B_0\hat{f}_i \\
         & - \|h_x(x)\|\left(\|H(s)\|_{\mathcal{L}_1} + \|L_0C_0G(s)\|_{\mathcal{L}_1} \right)\\
         & \cdot B_0 \gamma(\omega_o, T)-\|h_x(x)-h_{x}(\hat{x})\|\phi.
    \end{split}
\end{equation}

If the system states are not available for CBF-QP formulation, the following theorem shows the efficacy of the ESOR-QP based on state estimation. 
\begin{thm}
    For any \(t>0\), the condition
        \begin{equation}\label{eq_35}
            \sup\limits_{u\in U}\Psi_h(t,\hat{x},u)\geq -\beta(h(\hat{x}))
        \end{equation}
        is a sufficient condition for (\ref{eq_8}), and also a necessary condition for (\ref{eq_8}) when \(T\rightarrow 0\) and \(0\leq \omega_o<1\) for \(t>0\). 
\end{thm}

Our ESOR-QP is given as follows: 
\begin{align}
    u^* (\hat{x}) & = \operatorname*{arg\,min}_{u\in U} \|u-k(\hat{x})\|^2   \notag \\
    \text{s.t.} 
    \quad & \Psi_h(t,\hat{x},u) +\beta (h(\hat{x})) >0, \label{eq_37}
\end{align}
where \(\Psi_h(t,\hat{x},u)\) is defined in (\(\ref{eq_33}\)). To be less conservative, \(\|H(s)\|_{\mathcal{L}_1} + \|L_0C_0G(s)\|_{\mathcal{L}_1}\) and \(\|h_x(x)-h_{x}(\hat{x})\|\) in (\ref{eq_33}) can be assumed to be one and zero, respectively, in the steady state because the state estimation from the observer has converged to its actual value.


\section{Simulation Results}
\label{sec:simulation}

\subsection{Adaptive Cruise Control}
The proposed ESOR-QP is applied to a typical ACC problem \cite{xu2015robustness, zhao2020adaptive} in this subsection. The ego car equipped with an ACC system cruises at a desired speed \(v_d\) while maintaining a safe distance from the lead car. The speeds of the lead car and the ego car are \(v_l\) and \(v_f\), respectively. The distance \(D\) between two cars is measured by a radar.


By defining the system state as \(x=[v_f, D]^T\), the dynamic system is described as:
\begin{equation}\label{eq_38}
\begin{split}
    \begin{bmatrix}
        \dot{v}_f\\
        \dot{D}
    \end{bmatrix}&=
    \underbrace{\begin{bmatrix}
        -F_r(v_f)/m\\
        -v_f
    \end{bmatrix}}_{f(x)}+
    \underbrace{\begin{bmatrix}
        1/m\\
        0
    \end{bmatrix}}_{g(x)}u+
    \underbrace{\begin{bmatrix}
        d_0\\
        v_l
    \end{bmatrix}}_{d(t)}\\
    y&=\begin{bmatrix}
        v_f & D
    \end{bmatrix}^T
    \end{split}
\end{equation}
where \(m\) is the mass of the following car, \(u\) is the control input, \(F_r(v_f)=f_0+f_1v_f+f_2v_f^2\) is the aerodynamic drag term with coefficients \(f_0\), \(f_1\), and \(f_2\), \(d_0(t)\) is an external disturbance. \(d(t)\) includes \(d_0(t)\) and \(v_l\) because they both are unknown to the ego car.

Safety (safe distance keeping) and stability (desired velocity tracking) are two major objectives of the ACC controller design. The safety constraint of CBF is established based upon a safety function \(h\), which is usually a distance function for collision avoidance. In this particular ACC problem, \(h=D-\tau_d v_f\), where \(\tau_d\) is called \emph{headway}. The stability constraint is implemented using a CLF with an energy-like function \(V=(v_f-v_d)^2\) \cite{ames2016control}. The values of the parameters can be found in \cite{zhao2020adaptive}. The disturbance \(d_0=0.2g\sin{(2\pi t/10)}\) is imposed on the ego car.


Based on the discussion in Section II-B, the ACC system can be divided into the following two subsystems:
\begin{equation}\label{eq_40}
    \begin{cases}
        \dot{v}_f=-\frac{1}{m}F_r(v_f)+\frac{1}{m}u+d_0\\
        y_1 = v_f,
    \end{cases}
\end{equation}
\begin{equation}\label{eq_41}
    \begin{cases}
        \dot{D}=-v_f+v_l\\
        y_2=D.
    \end{cases}
\end{equation}
Since there are two measurements, two disturbances can be estimated. Two ESOs can be designed as follows:
\begin{equation}\label{eq_42}
\small
    \begin{bmatrix}
        \dot{\hat{v}}_f\\
        \dot{\hat{d}}_0
    \end{bmatrix}=A_1\begin{bmatrix}
        \hat{v}_f\\
        \hat{d}_0
    \end{bmatrix}+\begin{bmatrix}
        \frac{1}{m}\\
        0
    \end{bmatrix}u+\begin{bmatrix}
        -\frac{1}{m}\\
        0
    \end{bmatrix}F_r(y_1)+L_1\left(y_1-C_1\begin{bmatrix}
        \hat{v}_f\\
        \hat{d}_0
    \end{bmatrix}\right),
\end{equation}
\begin{equation}\label{eq_43}
\small
    \begin{bmatrix}
        \dot{\hat{D}}\\
        \dot{\hat{v}}_l
    \end{bmatrix}=A_2\begin{bmatrix}
        \hat{D}\\
        \hat{v}_l
    \end{bmatrix}+\begin{bmatrix}
        -1\\
        0
    \end{bmatrix}y_1+L_2\left(y_2-C_2\begin{bmatrix}
        \hat{D}\\
        \hat{v}_l
    \end{bmatrix}\right)
\end{equation}
where \(A_1=A_2=\begin{bmatrix}
    0 & 1\\
    0 & 0
\end{bmatrix}\), \(C_1=C_2=[1, 0]\), and \(L_1\) and \(L_2\) are observer gains for those two ESOs, determined by the observer bandwidth. The observer bandwidth of ESO \(\omega_o =20\) rad/s and the observer gain of DOB \(k_b=10\) are used in the simulation. 

To compute the estimation error bounds, the \(l_{f1}\) for \(d_0\) is \(0.2g(2\pi)/10\), and the \(l_{f2}\) for \(v_l\) is \(4\). Then \(\gamma_1(\omega_{o1},T)\) and \(\gamma_2(\omega_{o2},T)\) are only related to the observer bandwidth of each ESO and estimation sample time \(T\), where the sum of \(p(k)\) can be calculated numerically, and \(T=1\times 10^{-4}\) s for good accuracy. Note that the observer bandwidth of each ESO should be converted to the discrete-time domain to compute the estimation error bound.

\begin{figure}[htbp]
    \centering
    \includegraphics[width=0.9\linewidth]{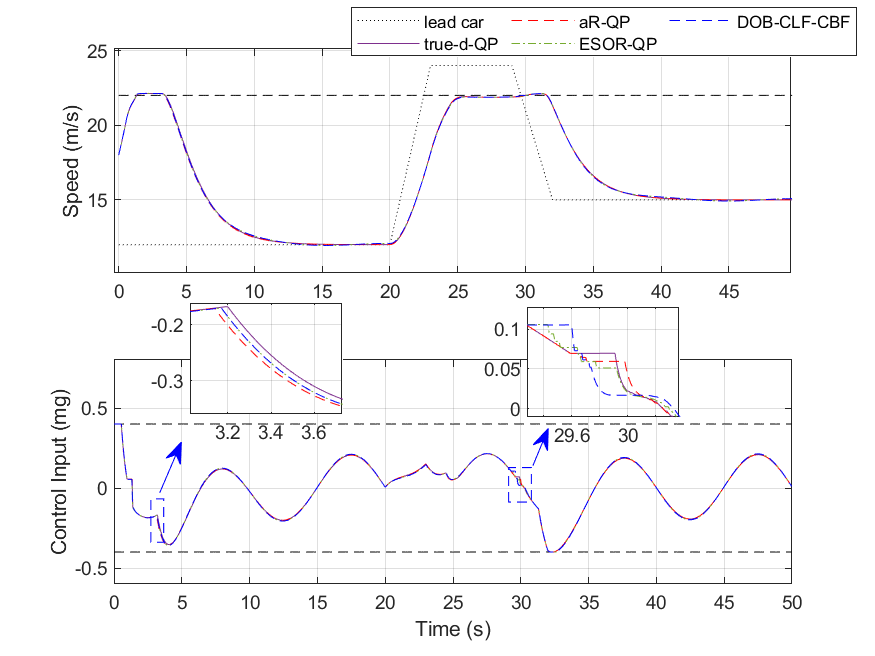}
    \caption{Comparison of QP with true \(d\) (called true-d-QP), aR-QP, ESOR-QP, and DOB-CLF-CBF-QP. (Top) speed of the lead car and following car with the desired speed \(v_d\) denoted by the black dotted line. (Bottom) control input as fractions of \(g\) with input limits denoted by black dashed lines.}
    \label{fig1}
\end{figure}

\begin{figure}[htbp]
    \centering
    \includegraphics[width=0.9\linewidth]{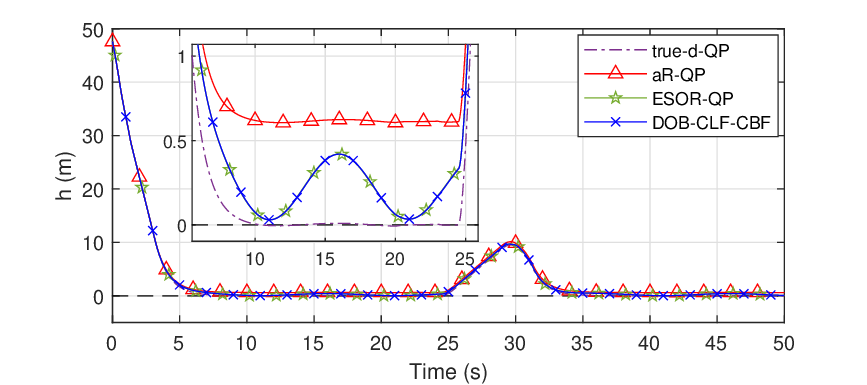}
    \caption{Comparison of safety function values between QP with true \(d\), aR-QP, ESOR-QP, and DOB-CLF-CBF-QP.}
    \label{fig2}
\end{figure}

To compare with \cite{zhao2020adaptive} and \cite{dacs2022robust}, we assume that the speed of the lead car \(v_l\) is known (directly measurable), i.e., we only need one ESO in (\ref{eq_42}). The profiles of the speed and the control input of the following car by using aR-QP, ESOR-QP and DOB-CLF-CBF-QP are shown in Fig. \ref{fig1}. The standard CLF-CBF-QP controller using the true uncertainty is also in Fig. \ref{fig1} as the perfect performance. Their trajectories of the safety function are shown in Fig. \ref{fig2}.

As shown in Fig. \ref{fig1}, since the disturbances are estimated and compensated in aR-QP, ESOR-QP, and DOB-CLF-CBF-QP, their control performances are close to the ideal case of known uncertainty. For the safety performance shown in Fig. \ref{fig2}, the aR-QP is the most conservative among them, and ESOR-QP and DOB-CLF-CBF-QP have the same safety performance as their estimation error bounds are both derived from their observer error dynamics. 


\subsection{Segway Robot Control}
The ESOR-QP is applied to a Segway self-balancing scooter with a high relative-degree CBF, whose dynamics and parameters are from \cite{molnar2022safety}. The wheel center position \(p\) and pitch angle \(\varphi\) are measured. The dynamic model is: 
\begin{equation}\label{eq_segway1}
    \underbrace{\begin{bmatrix}
        \dot{p}\\
        \dot{\varphi}\\
        \dot{\upsilon}\\
        \dot{\omega}
    \end{bmatrix}}_{\dot{x}}=\underbrace{\begin{bmatrix}
        \upsilon \\
        \omega \\
        f_{\upsilon}(\varphi, \upsilon, \omega)\\
        f_{\omega}(\varphi, \upsilon, \omega)
    \end{bmatrix}}_{f(x)}+\underbrace{\begin{bmatrix}
        0\\
        0\\
        g_{\upsilon}(\varphi)\\
        g_{\omega}(\varphi)
    \end{bmatrix}}_{g(x)}u+\underbrace{\begin{bmatrix}
        0\\
        0\\
        d_1\\
        d_2
    \end{bmatrix}}_{d(x)}, 
\end{equation}
where \(\upsilon\) and \(\omega\) are the velocities of \(p\) and \(\varphi\), respectively, \(d_1\) and \(d_2\) are two separate external disturbances, and other parameters can be found in \cite{molnar2022safety}. The safety constraint of CBF is chosen as \(h=\pi/10-\varphi^2\) to keep the Segway upright. The following nominal control law is used to track the desired wheel center position \(p_d\):
\begin{equation}\label{eq_segway2}
    k(x)=K_p(p-p_d)+K_{\upsilon}\upsilon+K_{\varphi}\varphi+K_{\omega}\omega 
\end{equation}
where gains are tuned to be \(K_p=4\) V/m, \(K_{\upsilon}=8\) Vs/m, \(K_{\varphi}=40\) V/rad, \(K_{\omega}=10\) Vs/rad, and the desired wheel position \(p_d=1\) m. The disturbances are set to \(d_1(t)=2\sin(2\pi t/10)\) and \(d_2(t)=2\cos(2\pi t/10)\). 

Based on the discussion in Section II-B, the Segway platform, as a signle-input multi-output system, can be divided into the following two subsystems:
\begin{equation}\label{eq_segway3}
    \begin{cases}
        \begin{bmatrix}
            \dot{p}\\
            \dot{\upsilon}
        \end{bmatrix}=\begin{bmatrix}
            0 & 1\\
            0 & 0
        \end{bmatrix}\begin{bmatrix}
            p\\
            v
        \end{bmatrix}+\begin{bmatrix}
            0\\
            g_\upsilon
        \end{bmatrix}u+\begin{bmatrix}
            0\\
            f_\upsilon
        \end{bmatrix}+\begin{bmatrix}
            0\\
            d_1
        \end{bmatrix}\\
        y_1 = p,
    \end{cases}
\end{equation}
\begin{equation}\label{eq_segway4}
    \begin{cases}
        \begin{bmatrix}
            \dot{\varphi}\\
            \dot{\omega}
        \end{bmatrix}=\begin{bmatrix}
            0 & 1\\
            0 & 0
        \end{bmatrix}\begin{bmatrix}
            \varphi\\
            \omega
        \end{bmatrix}+\begin{bmatrix}
            0\\
            g_\omega
        \end{bmatrix}u+\begin{bmatrix}
            0\\
            f_\omega
        \end{bmatrix}+\begin{bmatrix}
            0\\
            d_2
        \end{bmatrix}\\
        y_2=\varphi.
    \end{cases}
\end{equation}
Therefore, two ESOs like (\ref{eq_42}) and (\ref{eq_43}) can be designed to estimate the states and disturbances. The disturbances \(d_1\) and \(d_2\) can be directly estimated by the two ESOs, while DOB proposed in \cite{dacs2022robust} can only estimate the effect of the disturbances on the time derivative of CBF \(h\). The disturbance estimation using DOB-CBF-QP is given by
\begin{equation}\label{eq_segway5}
    \underbrace{\ddot{h}(x,u,d)}_{\dot{h}^{'}(x,u,d)} = \underbrace{L_f^2 h(x)+L_gL_f h(x) u}_{a_e(x,u)}+ \underbrace{L_g L_f h(x) d}_{b_e(x,d)}.
\end{equation}
Therefore, the effect of the disturbances on the CBF using DOB-CBF-QP is \(b_e(x,d)=-2\varphi d_2\). If \(b_e(x,d)\) is differentiable in \(t\), let \(|\dot{b}_e(x,d)|\leq b_h\). The estimation error bound of \(\hat{b}_e\) in steady state is equal to \(b_h/k_b\), where \(k_b\) is the observer gain of DOB. The estimation error bound for Segway platform using DOB is conservative because \(b_e\) depends on not only the disturbance $d_2$ but also the state \(\varphi\), which results in amplified error bound \(b_h\) due to the change of \(\varphi\). In contrast, the estimation error bound using ESO is only related to the time derivative bound of \(d_2\).

\begin{figure}[htbp]
    \centering
    \includegraphics[width=0.9\linewidth]{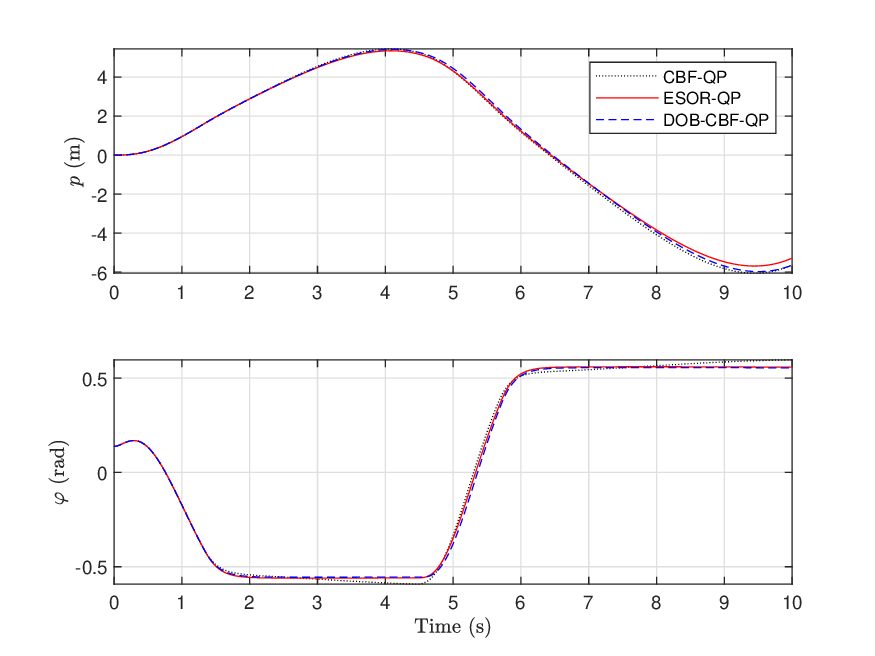}
    \caption{Comparison of QP with nominal CBF-QP, ESOR-QP, and DOB-CBF-QP. (Top) Trajectories of \(p\) [m]. (Bottom) Trajectories of \(\varphi\) [rad].}
    \label{fig4}
\end{figure}

\begin{figure}[htbp]
    \centering
    \includegraphics[width=0.9\linewidth]{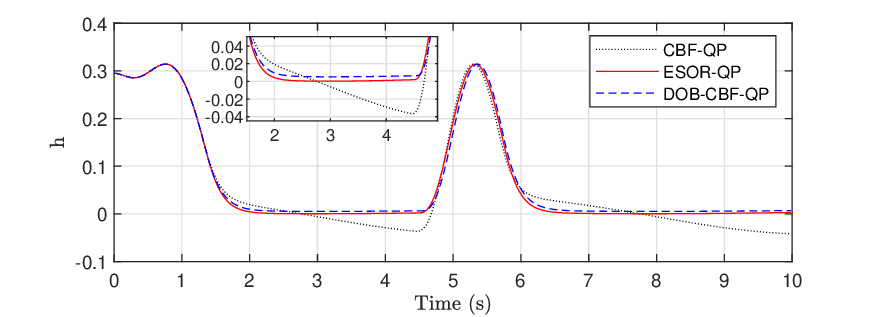}
    \caption{Comparison of safety function values between QP with nominal CBF-QP, ESOR-QP, and DOB-CBF-QP.}
    \label{fig5}
\end{figure}

The observer bandwidth of ESO \(\omega_o =20\) rad/s and the observer gain of DOB \(k_b=10\) are used in the simulation. Fig. \ref{fig4} and Fig. \ref{fig5} show simulation results of ESOR-QP compared with nominal CBF-QP (without considering any uncertainty) and DOB-CBF-QP. Due to the external disturbances and safety constraints, the wheel center position \(p\) cannot converge to the desired value 1 m. To improve the tracking performance on \(p\), a new disturbance rejection controller rather than the state-feedback control law in (\ref{eq_segway2}) should be designed. As shown in Fig. \ref{fig5}, with the assistance of ESO and DOB, the Segway robot always stays within the safe set under the external disturbances. However, the control of DOB-CBF-QP is more conservative than that of ESOR-QP, see the safety function value of DOB-CBF-QP is higher than our approach. Remember that in the ACC example, these two approaches get the same results because the disturbance in CBF for DOB-CBF-QP is not dependent on its state in that particular system. 

\section{CONCLUSIONS}
\label{sec:conclusion}

This paper studies a novel controller design, called ESOR-QP, combining an ESO with a CBF for provably safe control of uncertain dynamics without access to full state. Compared with state-of-the-art observer-based CBF-QP, we succeeds in obtaining a tighter estimation error bound to mitigate overconservatism. The comparisons in a cruise control system and a Segway robot validate the efficacy of our approach. Our future work includes the improvement of the nominal control law to minimize the interventions of QP-CBF and the testing in more complex robotic systems.

\bibliographystyle{IEEEtran}
\bibliography{references}

\end{document}